\documentclass[reprint,superscriptaddress,amssymb,amsmath,aps,showpacs,10pt,longbibliography,prb]{revtex4-2}
\usepackage{graphicx}

\graphicspath{{figures/}}
\usepackage{color}
\usepackage[colorlinks,bookmarks=false,citecolor=darkblue,linkcolor=red,urlcolor=blue]{hyperref}

\definecolor{darkred}{rgb}{0.7,0.0,0.0}
\definecolor{darkblue}{rgb}{0,0.02,0.45}

\definecolor{darkgreen}{rgb}{0.02,0.45,0.0}

\begin{document}


\title{Pressure-tuned spin chains in brochantite, Cu$_4$SO$_4$(OH)$_6$}

\author{Victoria A. Ginga}
\email{victoria.ginga@uni-leipzig.de}
\affiliation{Felix Bloch Institute for Solid-State Physics, University of Leipzig, 04103 Leipzig, Germany}

\author{Bin Shen}
\affiliation{Experimental Physics VI, Center for Electronic Correlations and Magnetism,
University of Augsburg, 86159 Augsburg, Germany}

\author{Ece Uykur}
\affiliation{Helmholtz-Zentrum Dresden-Rossendorf, Inst Ion Beam Phys \& Mat Res, D-01328 Dresden, Germany}

\author{Nico Giordano}
\affiliation{Deutsches Elektronen-Synchrotron DESY, 22607 Hamburg, Germany}

\author{Philipp Gegenwart}
\affiliation{Experimental Physics VI, Center for Electronic Correlations and Magnetism,
University of Augsburg, 86159 Augsburg, Germany}

\author{Alexander A. Tsirlin}
\email{altsirlin@gmail.com}
\affiliation{Felix Bloch Institute for Solid-State Physics, University of Leipzig, 04103 Leipzig, Germany}


\begin{abstract}
Using high-pressure single-crystal x-ray diffraction combined with thermodynamic measurements and density-functional calculations, we uncover the microscopic magnetic model of the mineral brochantite, Cu$_4$SO$_4$(OH)$_6$, and its evolution upon compression. The formation of antiferromagnetic spin chains with the effective intrachain coupling of $J\simeq 100$\,K is attributed to the occurrence of longer Cu--Cu distances and larger Cu--O--Cu bond angles between the structural chains within the layers of the brochantite structure. These zigzag spin chains are additionally stabilized by ferromagnetic couplings $J_2$ between second neighbors and moderately frustrated by several antiferromagnetic couplings that manifest themselves in the reduced N\'eel temperature of the material. Pressure tuning of the brochantite structure keeps its monoclinic symmetry unchanged and leads to the growth of antiferromagnetic $J$ with the rate of 3.2\,K/GPa, although this trend is primarily caused by the enhanced ferromagnetic couplings $J_2$. Our results show that the nature of magnetic couplings in brochantite and in other layered Cu$^{2+}$ minerals is controlled by the size of the lattice translation along their structural chains and by the extent of the layer buckling. 
\end{abstract}

\maketitle


\section{Introduction}
The field of mineralogy has been pivotal in developing functional materials with distinct chemical and physical properties. Minerals and their synthetic analogues with complex interaction geometries of transition-metal ions show intriguing magnetic properties. For example, copper minerals with their spin-$\frac12$ Cu$^{2+}$ ions are of significant interest for studying collective phenomena in quantum and frustrated magnets~\cite{inosov2018,zhang2020,smaha2020,fu2021,wang2022}.

The diversity of copper minerals goes hand in hand with their structural complexity. In these materials, leading magnetic interactions rarely follow the shortest \mbox{Cu--Cu} distances~\cite{jeschke2011,fujihala2022}, and even dimensionality of the spin lattice may deviate from the dimensionality of the underlying structural units~\cite{lebernegg2013a,janson2016,ginga2024}. Tailoring such materials toward the desired quantum regime requires a thorough microscopic understanding of individual magnetic interactions for their eventual control by external stimuli, such as pressure and strain.

\begin{figure*}[t]
  \centering
  \includegraphics[width=1\textwidth]{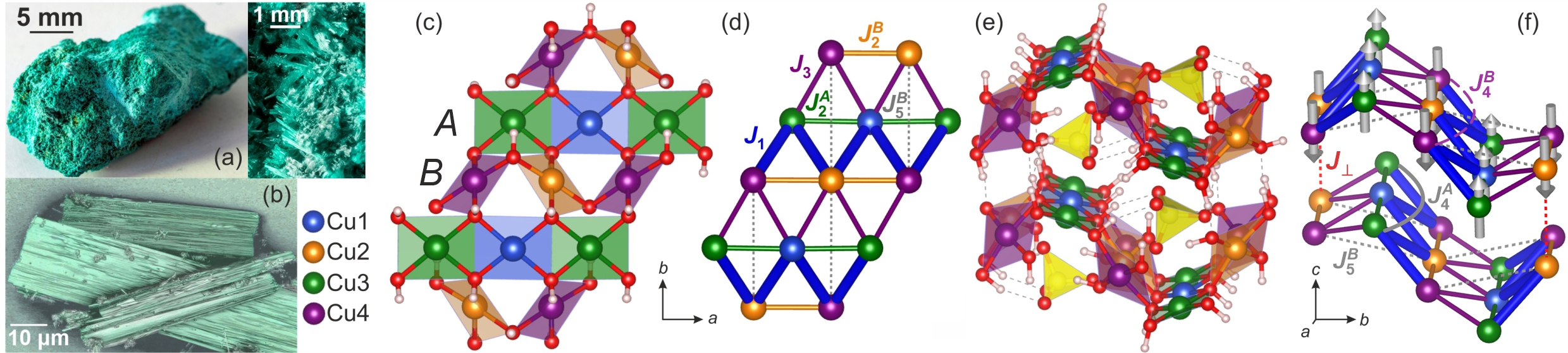}
    \caption{(a) Natural brochantite specimen from Bisbee, Arizona, USA, with flat prismatic plate crystals. (b) Optical and laser image of brochantite crystal aggregates taken using a Keyence VK-X200 microscope. The ambient-pressure crystal structure and the microscopic magnetic model of brochantite along $c$ (c,d) and $a$ (e,f) axes. In panel (f), experimental magnetic structure~\cite{vilminot2006,nikitin2023} is also shown for one of the layers.}
    \label{fig1}
\end{figure*}

Here, we consider brochantite, Cu$_4$SO$_4$(OH)$_6$, one of the ancient green pigments~\cite{valadas2015} and an integral component of copper patina~\cite{livingston1991}. From the magnetism perspective, it serves as a prototype of a large class of layered copper minerals. Its structural layers formed by Cu and O atoms comprise two ubiquitous units of cuprate structures, the chains of the CuO$_4$ plaquettes connected by edge-sharing (A) and corner-sharing (B), as shown in Fig.~\ref{fig1}c,e. Such structural chains are building blocks for a large variety of Cu-based materials where they may form not only infinite layers as in brochantite, but also isolated chains~\cite{moeller2009,caslin2014}, multi-chain ribbons~\cite{lebernegg2017}, and complex three-dimensional frameworks~\cite{heinze2021}. Understanding magnetic interactions within and between these chains, as well as exploring their pressure tunability are both important tasks for elucidating the magnetic behavior of a large class of prospective quantum magnets.

Previous studies of brochantite revealed its low-dimensional magnetic behavior manifested by a broad maximum in the magnetic susceptibility around 60\,K followed by the magnetic ordering at $T_N\simeq 6$\,K~\cite{vilminot2006,peets2025}. Collinear magnetic structure determined by neutron diffraction features ferromagnetic (FM) order along the structural chains and antiferromagnetic (AFM) order between the chains, as shown in Fig.~\ref{fig1}f~\cite{vilminot2006,nikitin2023}. This remarkably simple ground state was elucidated in an inelastic neutron scattering study~\cite{nikitin2023} that revealed the zigzag geometry of AFM spin chains spanning the adjacent structural chains A and B ($J_1$ in Fig.~\ref{fig1}d). Interestingly, all other magnetic couplings, including the couplings along the structural chains, were identified as negligible in that work, whereas an anisotropic exchange term was invoked to explain the observed magnon gap and helical spin dynamics~\cite{nikitin2023}. Zn-substituted brochantite, ZnCu$_3$(OH)$_6$SO$_4$, received further attention as a spin-liquid candidate with the kagome-like network of the Cu$^{2+}$ ions~\cite{li2014,gomilsek2016,gomilsek2017} and the Kondo screening effect~\cite{gomilsek2019}.

In the following, we use \textit{ab initio} calculations combined with the experimental probes to determine the microscopic magnetic model of brochantite and its evolution under pressure. We seek to understand how the complex structure of brochantite comes down to the simple model of AFM spin chains. We further explore pressure-induced changes in the crystal structure and magnetism of brochantite and establish quantitative links between the evolution of individual exchange couplings and the underlying structural geometry. Our work sets a benchmark for pressure studies of Cu-based quantum magnets and helps in rationalizing the magnetic behavior of structurally complex Cu$^{2+}$ compounds.

\section{Methods}

Experiments were performed on natural thin prismatic crystals of brochantite obtained from Bisbee, Arizona, USA (Fig. \ref{fig1}a). Some of the crystals were ground into fine powder for collecting high-resolution ambient-pressure x-ray diffraction (XRD) data at 80\,K and 298\,K at the ID22 beamline of the European Synchrotron Radiation Facility (ESRF) in Grenoble, France using a wavelength of 0.35433\,\r A and the multi-analyzer detector setup~\cite{fitch2023}. Additionally, selected crystals were studied at ambient pressure and room temperature on a STOE Stadivari diffractometer equipped with a GeniX 3D Cu-HF (Xenocx) microfocus X-ray source and a hybrid pixel detector Pilatus 300K (Dectris). Data processing was carried out with the STOE X-Area software package~\cite{StoeCie2019}.


Thermodynamic properties at ambient pressure were studied on the ground powder sample. Magnetization was measured on a Quantum Design MPMS-XL SQUID magnetometer over a temperature range of $2-280$ K, with the applied fields of 0.01\,T, 0.1\,T, 1\,T and 5\,T. The heat capacity ($C_p$) was measured on a pressed pellet over a temperature range from 1.8\,K to 200\,K using the Physical Property Measurement System (PPMS) from Quantum Design.

High-pressure single-crystal XRD data were collected at room temperature at the P02.2 beamline~\cite{p02-2} of the PETRA\,III synchrotron (DESY, Hamburg, Germany) using x-rays with the wavelength of 0.2910\,\r A and a Perkin Elmer XRD\,1621 detector. Diffraction images were obtained by a $\varphi$-rotation of the pressure cell between $-30$ and $+30^{\circ}$ with a step of $0.5^{\circ}$ and integrated using \texttt{CrysAlisPro} software~\cite{crysalispro}. \texttt{Jana2006}~\cite{jana2006} was used for all structure refinements. A single crystal, measuring approximately $30\times 30\times 10$\,$\mu$m$^3$, was placed inside a diamond anvil cell (DAC), comprising Type Ia Boehler-Almax cut diamonds with a culet diameter of 300\,$\mu$m. Neon gas was used as the pressure-transmitting medium. Pressure was determined through ruby luminescence measurements~\cite{mao1986}. 

Magnetization measurements under pressure were performed in a CuBe cell with a 900 \,$\mu$m culet diameter using the MPMS\,3 magnetometer from Quantum Design following the procedure described in Ref.~\cite{shen2021}. Daphne oil 7373 was used as pressure-transmitting medium. The measurements were conducted on a collection of randomly oriented small crystals.

Magnetic couplings in brochantite were evaluated by density-functional theory (DFT) band-structure calculations performed in the \texttt{VASP} code~\cite{vasp1,vasp2} using the Perdew-Burke-Ernzerhof type of the exchange-correlation potential~\cite{pbe96}. Experimental lattice parameters and atomic positions for Cu, S, and O were used in all calculations. Hydrogen positions were optimized by DFT because of the lower sensitivity of XRD to hydrogen atoms. The accurate location of hydrogen atoms in the structure is known to be crucial for the correct evaluation of magnetic couplings~\cite{janson2008,lebernegg2013b}. The coupling parameters of the Heisenberg Hamiltonian,
\begin{equation}
 \mathcal H=\sum_{\langle ij\rangle} J_{ij}\mathbf S_i\mathbf S_j
\end{equation}
where $S=1/2$ and the summation is over bonds, were obtained by a mapping procedure~\cite{xiang2011} using total energies from DFT+$U$ calculations with $U_d=9.5$\,eV and $J_d=1$\,eV for the Cu $3d$ shell and the double-counting correction in the atomic limit~\cite{nekrasova2020,panther2023}. Additionally, we performed a tight-binding analysis of the band structure calculated in \texttt{FPLO}~\cite{fplo} for a cross-check of the DFT+$U$ results, as explained in Sec.~\ref{sec:ambient} below.


\section{Ambient-pressure behavior}
\label{sec:ambient}

\begin{table*}
\caption{\label{tab:exchange}
Exchange couplings calculated for the $P2_1/a$~\cite{helliwell1997} and $P2_1/n$~\cite{merlino2003} polymorphs of brochantite at ambient pressure, with all hydrogen positions relaxed. The $d_i$ and $\alpha_i$ stand for the Cu--Cu distances and the respective bond angles (Cu--O--Cu for $J_1-J_3$ and Cu--O--O in the case of $J_4$, $J_5$, and $J_{\perp}$). The $J_i$ values are obtained from the DFT+$U$ mapping analysis, whereas $t_i$ are the corresponding hoppings in the PBE band structure. The couplings are split into groups based on the similarity of the Cu--Cu distances and interaction geometries, with an average value calculated for each group and given in bold. The notation of the Cu atoms is shown in Fig.~\ref{fig1}.
}
\begin{ruledtabular}
\begin{tabular}{cc@{\hspace{2cm}}ccrr@{\hspace{2cm}}ccrr}
 & & $P2_1/a$ & & & & $P2_1/n$ & & & \smallskip\\
 & & $d_i$ (\r A) & $\alpha_i$ (deg) & $t_i$ (meV) & $J_i$ (K) & $d_i$ (\r A) & $\alpha_i$ (deg) & $t_i$ (meV) & $J_i$ (K) \medskip\\
 $J_1$ & Cu1--Cu2 & 3.561 & 121.3 & 124    &  88   &  3.557 & 122.2 & 122    & 107 \\
       & Cu1--Cu4 & 3.533 & 124.4 & $-132$ & 121   &  3.522 & 122.4 & 125    & 114 \\
       & Cu2--Cu3 & 3.561 & 119.9 & 122    &  88   &  3.573 & 121.2 & $-112$ & 72  \\
       & Cu3--Cu4 & 3.518 & 121.7 & $-117$ &  91   &  3.529 & 121.3 & $-118$ & 95  \\
            &     &     &   & &  \textbf{97}   &        &  &        & \textbf{97}\medskip\\
$J_2^A$ & Cu1--Cu3 & 2.996 & 97.7 &  $-79$ &   8   &  3.012 & 97.9 &  82   & 13  \\
        & Cu1--Cu3 & 3.020 & 98.4 & $-111$ &  36   &  3.015 & 97.7 &  96   & 22  \\
               &   &     &  & &  \textbf{22}   &        &  &       & \textbf{18}\medskip\\ 
$J_2^B$ & Cu2--Cu4 & 2.997 & 91.2 & $-45$  & $-29$ &  3.002 & 91.0 &  52   & $-28$ \\
        & Cu2--Cu4 & 3.019 & 91.9 & $-55$  & $-32$ &  3.025 & 92.2 &  53   & $-35$ \\
               &   &     &  & &  \textbf{--31} &        &  &       & \textbf{--32}\medskip\\
 $J_3$ & Cu1--Cu2 & 3.194 & 99.7  &  $-80$  &  4    &  3.243 & 97.0 &  78   &  18 \\
       & Cu1--Cu4 & 3.206 & 102.1 &    93   &  23   &  3.273 & 99.4 &  98   &  42 \\
			 & Cu2--Cu3 & 3.230 & 96.2 &  $-95$  &  28   &  3.204 & 99.7 & $-76$ &   9 \\
			 & Cu3--Cu4 & 3.265 & 99.2 &   106   &  56   &  3.219 & 101.1 & $-79$ &  12 \\
			       &    &     & & &  \textbf{28}    &        &  &       & \textbf{20}\medskip\\
$J_4^A$ & Cu1--Cu1 & 6.015 & 138.8 &  68    &  27   &  6.026 & 138.7 &  60   &  27\\
        & Cu3--Cu3 & 6.015 & 138.9 &  65    &  30   &  6.026 & 139.0 &  68   &  28\\
               &   &     & & &  \textbf{29}   &        & &       & \textbf{28}\medskip\\
$J_4^B$ & Cu2--Cu2 & 6.015 & &  34    &   1   &  6.026 & &  37   &  2 \\
        & Cu4--Cu4 & 6.015 & &  32    &   1   &  6.026 & &  32   &  1 \\
               &   &     & & &  \textbf{1}    &        & &       & \textbf{2}\medskip\\
$J_5^B$ & Cu2--Cu2 & 5.067 & 121.8 &  36   &  15    &  \\
        & Cu4--Cu4 & 5.043 & 122.2 & $-25$ &  15    &  \\
				& Cu2--Cu4 &       & &       &        &  5.067 & 122.2 &  34   &  15 \\
				& Cu2--Cu4 &       & &       &        &  5.075 & 122.2 &  30   &  15 \\
				&          &       &  & & \textbf{15}  &        &  &       & \textbf{15}\medskip\\
$J_{\perp}$ & Cu2--Cu2 & 5.366 & 116.7 &  42    &  10   &  5.377 & 119.3 &  38 & 9 \\
\end{tabular}
\end{ruledtabular}
\end{table*}

Natural brochantite features two monoclinic polymorphs with the different layer stacking~\cite{helliwell1997,merlino2003,mills2010}, whereas yet another, orthorhombic polymorph was recently stabilized in synthetic samples~\cite{peets2025}. The monoclinic polymorphs are MDO$_1$ ($P2_1/a$, $a\simeq 6.02$\,\r A, $b\simeq 9.85$\,\r A, $c\simeq 13.07$\,\r A, and $\beta\simeq 103.3^{\circ}$) and MDO$_2$ ($P2_1/n$, $a\simeq 6.02$\,\r A, $b\simeq 9.85$\,\r A, $c\simeq 12.72$\,\r A, and $\beta\simeq 90.0^{\circ}$). Our high-resolution powder XRD data~\cite{supplement} confirm the high crystallinity and excellent purity of the natural brochantite sample. These data can be described by either of the two monoclinic models, although $P2_1/n$ returns slightly lower refinement residuals of $R_1=0.040$ compared to $R_1=0.043$ for $P2_1/a$. Interestingly, single-crystal XRD data collected at ambient pressure are indicative of the $P2_1/a$ structure, but our high-pressure XRD data (Sec.~\ref{sec:highp}) systematically showed the $P2_1/n$ structure instead. One possible explanation for these observations is that the $P2_1/a$ structure transforms into the $P2_1/n$ one already at low pressures and even upon grinding, because the two polymorphs must be very close in energy. Below, we show that these two polymorphs are also very similar from the perspective of their magnetic models, and the occurrence of one or another monoclinic polymorph should have no significant influence on the magnetic behavior of brochantite.

The high-resolution powder XRD data collected at 80\,K showed only minor changes compared to room temperature. Therefore, we used the room-temperature crystallographic parameters in DFT calculations of the magnetic couplings. Both monoclinic polymorphs of brochantite feature four Cu sites with a plethora of nonequivalent exchange pathways. Fortunately, many of these pathways are similar to each other, making it convenient to distribute the couplings into groups (Table~\ref{tab:exchange}). The first group involves the Cu--Cu contacts of $3.5-3.6$\,\r A that correspond to antiferromagnetic (AFM) couplings on the order of 100\,K. These couplings are labeled $J_1$ following Ref.~\cite{nikitin2023}. They form zigzag spin chains that span the adjacent structural chains A and B (Fig. \ref{fig1}d,f). Despite the very short separations between these spin chains, the couplings between them given by $J_3$, with the Cu--Cu distances of about 3.2\,\r A, are relatively weak, leading to the quasi-1D magnetic topology.

Yet another group of couplings involves the shortest Cu--Cu contacts of 3.0\,\r A that occur within the A and B structural chains of brochantite. Following Ref.~\cite{nikitin2023}, we label these couplings as $J_2^A$ and $J_2^B$ because they describe second-neighbor interactions within the zigzag spin chains formed by $J_1$. One unexpected finding is that $J_2^A$ is AFM, whereas $J_2^B$ is FM following the different connectivity of the CuO$_4$ polyhedra (Fig.~\ref{fig1}c). 

We now assess the magnetic couplings in brochantite using the hopping parameters ($t_i$) as a gauge of AFM superexchange, $J_i^{\rm AFM}\sim t_i^2$. These hopping parameters elucidate the difference between $J_1$ and $J_3$. The couplings of the $J_3$ group feature smaller hoppings because of the lower Cu--O--Cu angles that average to $99^{\circ}$, in contrast to $122^{\circ}$ in the case of $J_1$. Additionally, the shorter Cu--Cu distances in the case of $J_3$ lead to larger FM contributions due to direct exchange. Therefore, the couplings of the $J_1$ group dominate, and the structural layers of brochantite split into spin chains. The effect of the FM contributions is even more pronounced in the case of $J_2^A$ and $J_2^B$ with their shorter Cu--Cu distances and Cu--O--Cu angles approaching $90^{\circ}$. The change from AFM $J_2^A$ to FM $J_2^B$ can be also ascribed to the reduction in the bond angle, in agreement with the Goodenough-Kanamori-Anderson rules. 

Another important finding is the occurrence of the long-range AFM couplings $J_4^A$ that connect second neighbors within the structural chains A. Such couplings are indeed very common in Cu chains with the edge-sharing geometry~\cite{kuzian2012}. On the other hand, they are usually suppressed in the Cu chains with the corner-sharing geometry~\cite{moeller2009} where the respective O--O distances are too large to facilitate the coupling. Indeed, we find a negligible $J_4^B$ in brochantite. 

Our microscopic magnetic models for the two monoclinic polymorphs of brochantite are not only qualitatively but also quantitatively similar. Therefore, we make no further distinction between these polymorphs and compare the model to the experimental data that are usually reported for the $P2_1/a$ structure. The collinear magnetic order in brochantite (Fig.~\ref{fig1}f) is well explained by the interplay of AFM $J_1$ and $J_3$ that cause FM order along the structural chains. This FM order is further stabilized by $J_2^B<0$ and weakly frustrated by $J_2^A>0$, but the main mechanism of frustration is the second-neighbor coupling $J_4^A>0$ that would lead to helical order if confronted with $J_2^A$ alone. However, in brochantite the FM order along the structural chains is stabilized by the much stronger coupling $J_1$, so the frustration by $J_4^A$ does not lead to a departure from collinearity. Our calculated exchange couplings are also in a favorable agreement with the inelastic neutron scattering (INS) results~\cite{nikitin2023}: compare $J_1^{\rm INS}=124$\,K with our value of 97\,K and $J_2^{\rm INS}\simeq 0$ with our $J_2\simeq -5$\,K obtained by averaging $J_2^A$ and $J_2^B$ from Table~\ref{tab:exchange}. 

Before turning to thermodynamic properties, we note in passing that our tight-binding analysis uncovers two further exchange couplings that are smaller than $J_1-J_4$ but still sizable. The AFM coupling $J_5^B$ connects chain B to another chain B, whereas $J_{\perp}$ is the AFM interlayer coupling that, surprisingly, occurs for every fourth Cu site only (Fig.~\ref{fig1}). Both $J_5^B$ and $J_{\perp}$ are long-range in nature and share the common mechanism of the \mbox{Cu--O--O--Cu} superexchange between two co-aligned CuO$_4$ plaquettes. The strength of these couplings, along with the other long-range coupling $J_4^A$, is controlled by the Cu--O--O angles. While the larger angle of about $139^{\circ}$ is observed for the stronger coupling $J_4^A$, the smaller angle of $117-119^{\circ}$ occurs in the case of the weaker coupling $J_{\perp}$. All further couplings, both within and between the structural layers of brochantite, are below 2\,K and thus negligible.

\begin{figure}[h]
  \centering
  \includegraphics[width=0.4\textwidth]{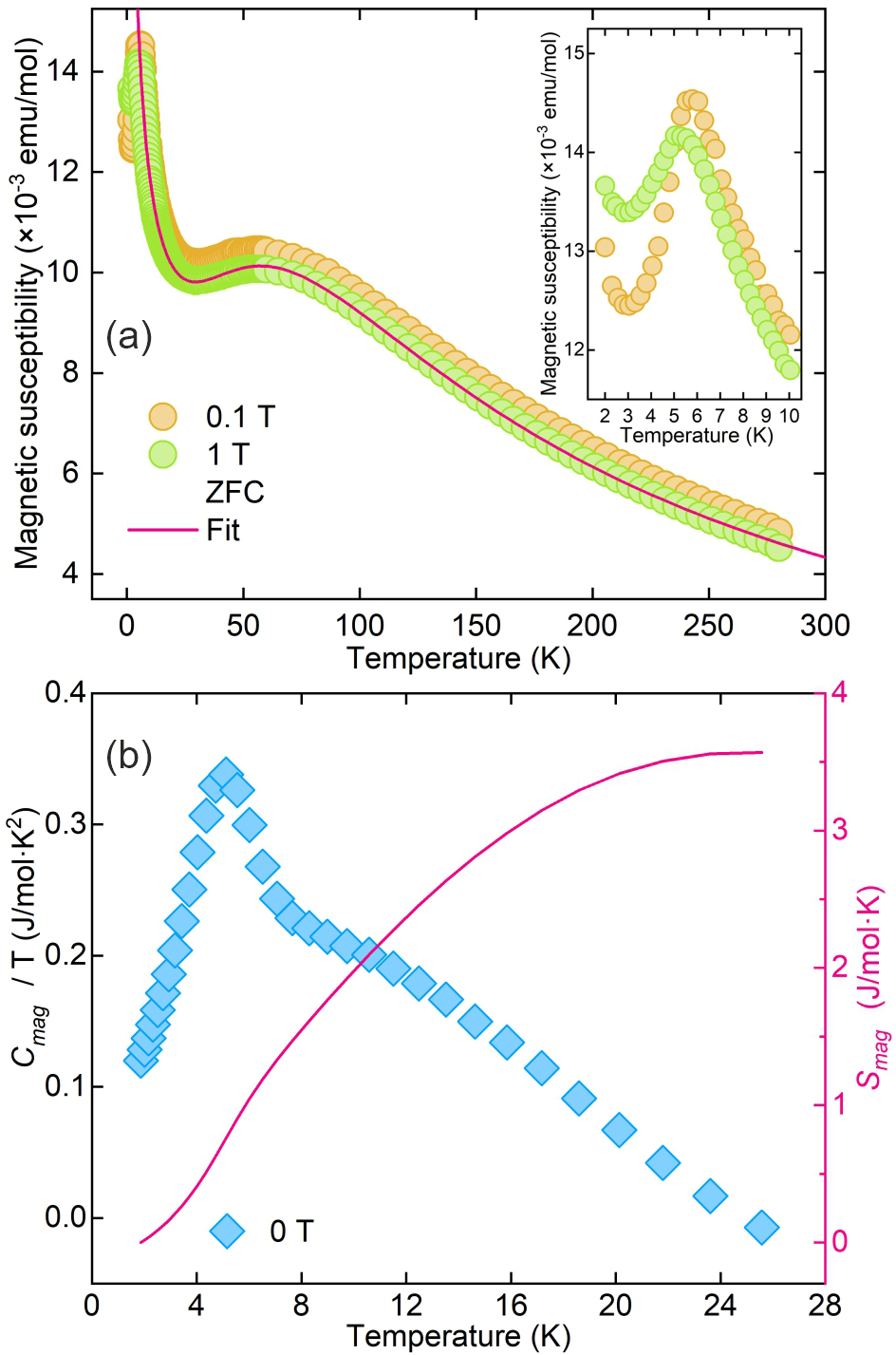}
    \caption{(a) Temperature-dependent magnetic susceptibility of brochantite measured in the applied fields of 0.1 T and 1 T. The inset shows magnetic transition in the low-temperature region. (b) $C_{\rm mag}/T$ of brochantite as a function of temperature at 0\,T and the calculated magnetic entropy $S_{\rm mag}$ in zero field.}
    \label{fig2}
\end{figure}

Fig. \ref{fig2}a shows the temperature-dependent magnetic susceptibility of our brochantite sample. Similar to the previous reports~\cite{vilminot2006,nikitin2023}, it shows a broad maximum around 60\,K followed by a minimum around 30\,K and an increase toward low temperatures. While the maximum arises from the spin chains pinpointed by our microscopic analysis, the origin of the upturn below 30\,K has not been revealed. Here, we model this upturn as a simple impurity contribution because its magnitude varies across the different samples reported in the literature. It is worth noting, though, that the synthetic sample of brochantite showed such an upturn for one field direction only~\cite{peets2025}. Its exact origin requires further investigation that goes beyond the scope of our present study. Here, we fit the experimental magnetic susceptibility with
\begin{equation}
 \chi(T)=\chi_0+\frac{C_{\rm imp}}{T-\theta_{\rm imp}}+\chi_{\rm ch}(T)
\label{eq:chi}
\end{equation}
where $\chi_{\rm ch}(T)$ is the magnetic susceptibility of a uniform spin-$\frac12$ chain given in Ref.~\cite{johnston2000}, $\chi_0$ is the temperature-independent contribution, and the second term stands for an impurity contribution described by the Curie-Weiss law. The fit of the experimental data down to 6\,K returns $J=104.3\pm 0.3$\,K in an excellent agreement with our \textit{ab initio} estimate of $J_1$ in Table~\ref{tab:exchange}. Further fit parameters are $g=2.17$, $C_{\rm imp}=0.060$\,emu\,K/mol, $\theta_{\rm imp}=-2.6$\,K, and $\chi_0=-7.3\times 10^{-4}$\,emu/mol. The $C_{\rm imp}$ value corresponds to 4\% of spin-$\frac12$ impurities per Cu$^{2+}$ ion. 

\begin{figure}[h]
  \centering
  \includegraphics[width=0.405\textwidth]{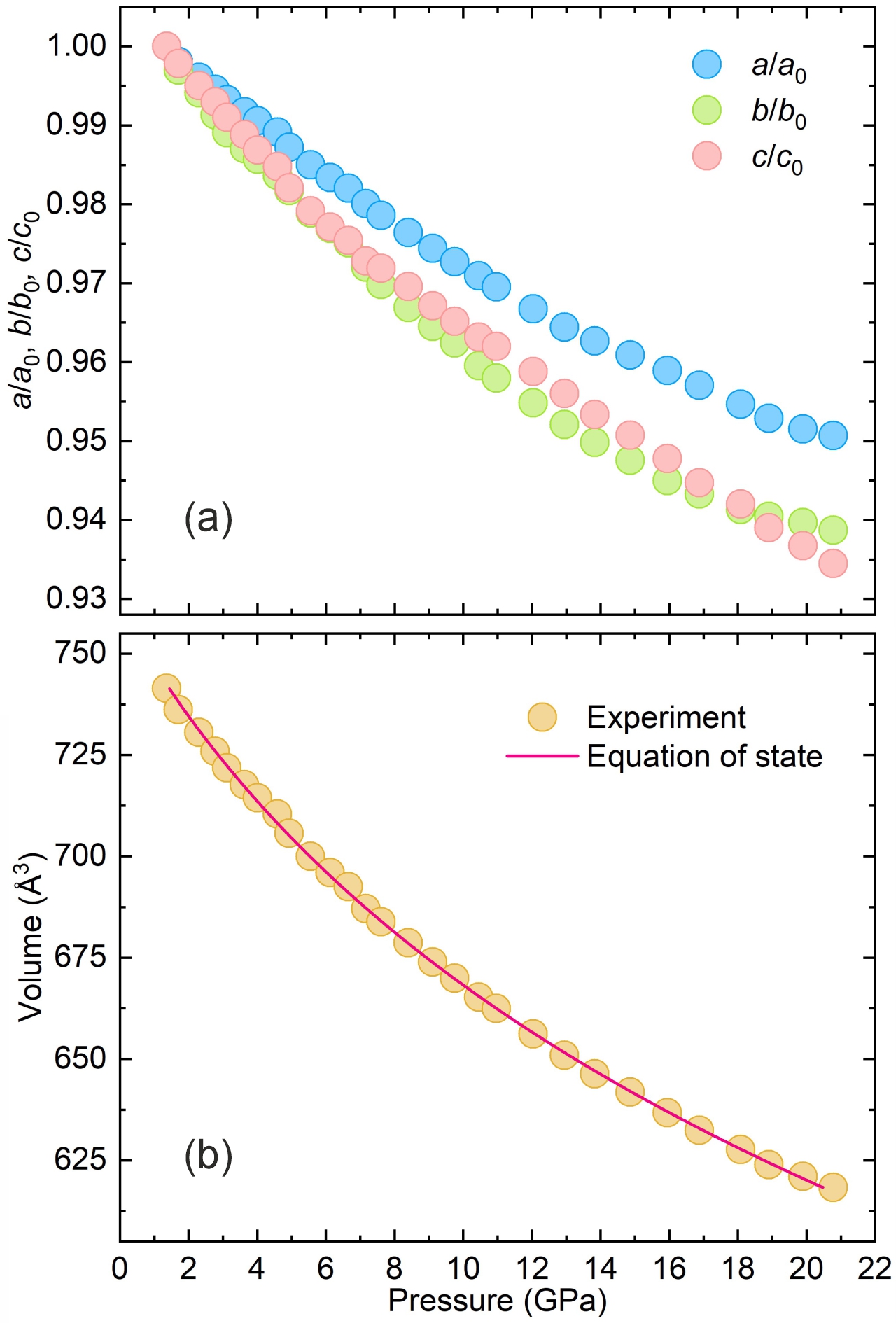}
    \caption{(a) The lattice constants $a$/$a_0$, $b$/$b_0$, $c$/$c_0$ as a function of pressure. (b) Pressure dependence of the unit-cell volume. The solid blue line is the fit to the equation of state as described in the text.}
    \label{fig3}
\end{figure}

Our susceptibility data measured at 0.1\,T also show a kink around $T_N=5.0$\,K that corresponds to the magnetic ordering. This $T_N$ value is somewhat lower than $6.0-6.5$\,K~\cite{nikitin2023,peets2025} and 7.5\,K~\cite{vilminot2006} observed in the previous studies, probably because of some sample dependence across the natural crystals of brochantite. It is worth noting that we do not find any signatures of an additional magnetic transition around 18\,K that has been seen in some of the brochantite samples~\cite{vilminot2006,bissengaliyeva2013}. Both magnetic susceptibility and specific heat data clearly show the transition at 5.0\,K only. This transition shifts toward lower temperature on increasing the field.

To estimate the magnetic contribution to the specific heat ($C_{\rm mag}$) and the corresponding magnetic entropy, we determined the phonon component ($C_{\rm ph}$) of the total heat capacity ($C_p$) by fitting the data in the range of $13-280$\,K~\cite{supplement} with a combination of three Debye functions and an $A/T^{2}$ term that stands for the high-temperature part of $C_{\rm mag}$:
\begin{equation}
 C_p(T)= A/T^{2} + \sum_{1}^{3}g_i\,C_{Deb, i}(\theta_{Deb, i}, T)
\label{eq:hc}
\end{equation}
where $g_i$ are pre-factors that determine the fractions of different contributions. Three Debye temperatures used for this fit were $D_1=190$\,K for 4 Cu atoms, $D_2=1280$\,K for one S atom, and $D_3=724$\,K for 10 oxygen atoms per formula unit. The light hydrogen atoms give rise to high-energy optical modes that can be neglected in the temperature range of our study. By subtracting the calculated phonon contribution from the experimental heat capacity data, the magnetic heat capacity was obtained (Fig. \ref{fig2}b). The magnetic entropy derived by integrating $C_{\rm mag}/T$ up to 25\,K is about 15\% of the total entropy $S_{\rm mag}=4\,R\ln 2\simeq 23.04$\,J\,mol$^{-1}$K$^{-1}$ expected for the four Cu$^{2+}$ ions (Fig. \ref{fig2}b), whereas only 4\% of the total magnetic entropy is released at $T<T_N$. This large reduction in the magnetic entropy confirms low-dimensional nature of brochantite magnetism. 

For a rough estimate of $T_N$, we calculate the averaged interchain coupling, $J_{\perp}^{\rm eff}=J_3+J_{\perp}/4=22$\,K where the factor of $\frac14$ is due to the fact that only one out of 4 Cu$^{2+}$ ions is coupled to the adjacent layer. The resulting $J_{\perp}^{\rm eff}/J_1\simeq 0.23$ would lead to $T_N/J_1\simeq 0.3$~\cite{yasuda2005} and $T_N\simeq 29$\,K, which is much larger than the experimental $T_N$ of $5-6$\,K. This comparison demonstrates that the spin lattice of brochantite must be frustrated, owing to the couplings $J_2^A$, $J_4^A$, and $J_5^B$ that are not satisfied by the experimental magnetic structure. The reduction in $T_N$ is then a combined effect of the low-dimensionality and frustration.

\section{Pressure evolution}
\label{sec:highp}

Our single-crystal XRD study revealed a gradual compression of the brochantite structure and the retention of the $P2_1/n$ symmetry up to at least 33\,GPa. In the following, we show the structural data up to 21\,GPa only, because technical difficulties in the data collection hindered complete structure refinements at higher pressures.


\begin{figure}
  \centering
  \includegraphics[width=0.4\textwidth]{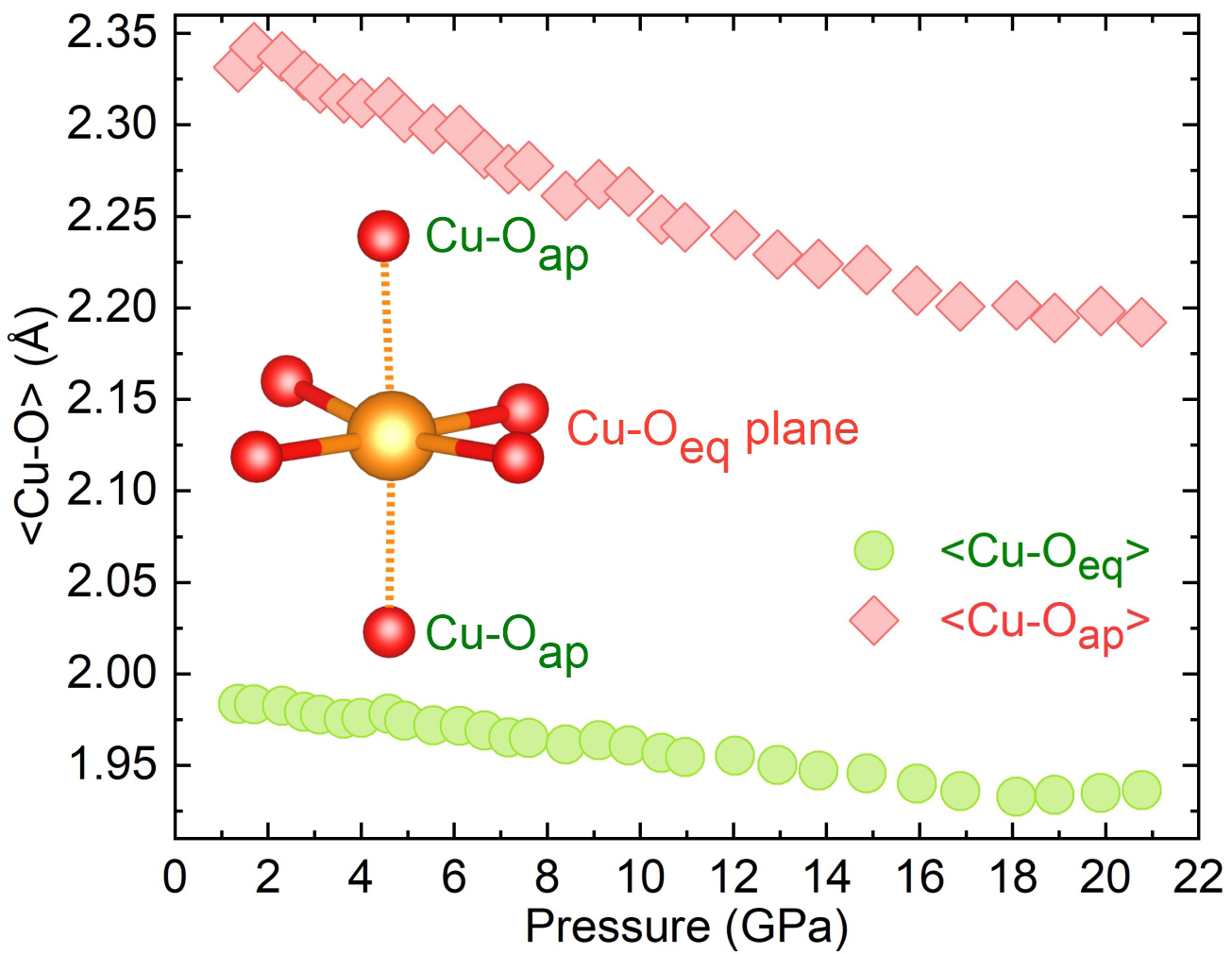}
    \caption{Pressure-dependent evolution of the average values of the Cu–O bond lengths in the crystal structure of brochantite. The insert shows the arrangement of apical and equatorial bonds in the coordination environment of the copper cation.}
    \label{fig4}
\end{figure}

The fit of the pressure-dependent volume with the Birch-Murnaghan equation of state (Fig. \ref{fig3}b) returns the equilibrium volume $V_0=761.075(1)$\,\r A$^3$/f.u., initial bulk modulus $B_0=50.3(1)$\,GPa, and pressure derivative of the bulk modulus $B_0'=6.62(2)$. Compressibility of brochantite is comparable to that of other hydroxy copper minerals, such as malachite with $B_0=48(4)$\,GPa~\cite{merlini2012}, 
botallackite with $B_0=51$\,GPa~\cite{zhao2020}, and azurite with $B_0=40(2)$\,GPa~\cite{xu2015}. Despite the layered nature of the brochantite structure, its compression is remarkably isotropic because of the multiple linkages between the layers. These linkages arise from the network of hydrogen bonds~\cite{supplement} and from the fact that Cu atoms of the adjacent layers are bonded via oxygen atoms of the SO$_4$ tetrahedra. Such Cu--O bonds are apical (perpendicular to the CuO$_4$ plaquettes) and longer than the equatorial bonds within the CuO$_4$ plane (Fig. \ref{fig4}). This characteristic Jahn-Teller distortion of Cu$^{2+}$ is preserved across the whole pressure range of our study, although the apical bonds show a much stronger compression, which is a common effect across the different Cu$^{2+}$ compounds~\cite{ruiz2011,caslin2016}.

The energy of hydrogen bonds in brochantite varies significantly, as evidenced by infrared and Raman spectroscopy~\cite{schmidt1993}. Our structural analysis of brochantite reveals two distinct groups of hydroxyl (-OH) sites with different hydrogen-bonding environments. Specifically, in two hydroxyl groups, (OH)$_2$ and (OH)$_6$, the hydrogen bond acceptors are other hydroxyl groups, forming interconnected OH$\,\cdots$OH networks. In the remaining four hydroxyl sites, the hydrogen bonds are instead donated to oxygen atoms of adjacent (SO$_{4}$)$^{2-}$ groups, stabilizing the sulfate framework. Under pressure, the most pronounced changes occur in the $D$--$H$ bonds of (OH)$_2$ and (OH)$_6$, where a neighboring hydroxyl acts as an acceptor. The longest $D$···$A$ distance, O6$\,\cdots$O5 (3.25\,\r A), corresponds to weaker bifurcated hydrogen bonds, yet this contact dramatically shortens to 2.62\,\r A under pressure~\cite{supplement}, suggesting a substantial strengthening of this interaction. In contrast, the shortest $D\cdots A$ distance, O1$\,\cdots$O8, which is 2.66\,\r A, decreases to 2.45\,\r A under pressure, reinforcing its role as a strong and directional hydrogen bond.

The observed pressure-induced enhancement of hydrogen bonding plays a crucial role in the structural stability of brochantite. Typically, layered structures are prone to an anisotropic compression that may even lead to an interlayer collapse. However, in brochantite, a complex network of hydrogen bonds reinforces the Cu-OH layers, mitigating this effect. Stronger OH-O(SO$_{4}$)$^{2-}$ interactions and the tightening of OH$\,\cdots$OH contacts under pressure contribute to an almost isotropic compression behavior, which is clearly seen in the very similar compression of the intralayer and interlayer lattice parameters $b$ and $c$, respectively (Fig.~\ref{fig4}). Such a weak elastic anisotropy is not typical for the layered hydroxy minerals. Additionally, the synergy between the Cu$^{2+}$ coordination and hydrogen bonding further stabilizes the structure by counteracting distortions within the sulfate-linked hydroxyl network. This highlights hydrogen bonding as a key factor in maintaining the structural integrity of brochantite across a wide pressure range.

\begin{figure*}[t]
  \centering
  \includegraphics[width=1\textwidth]{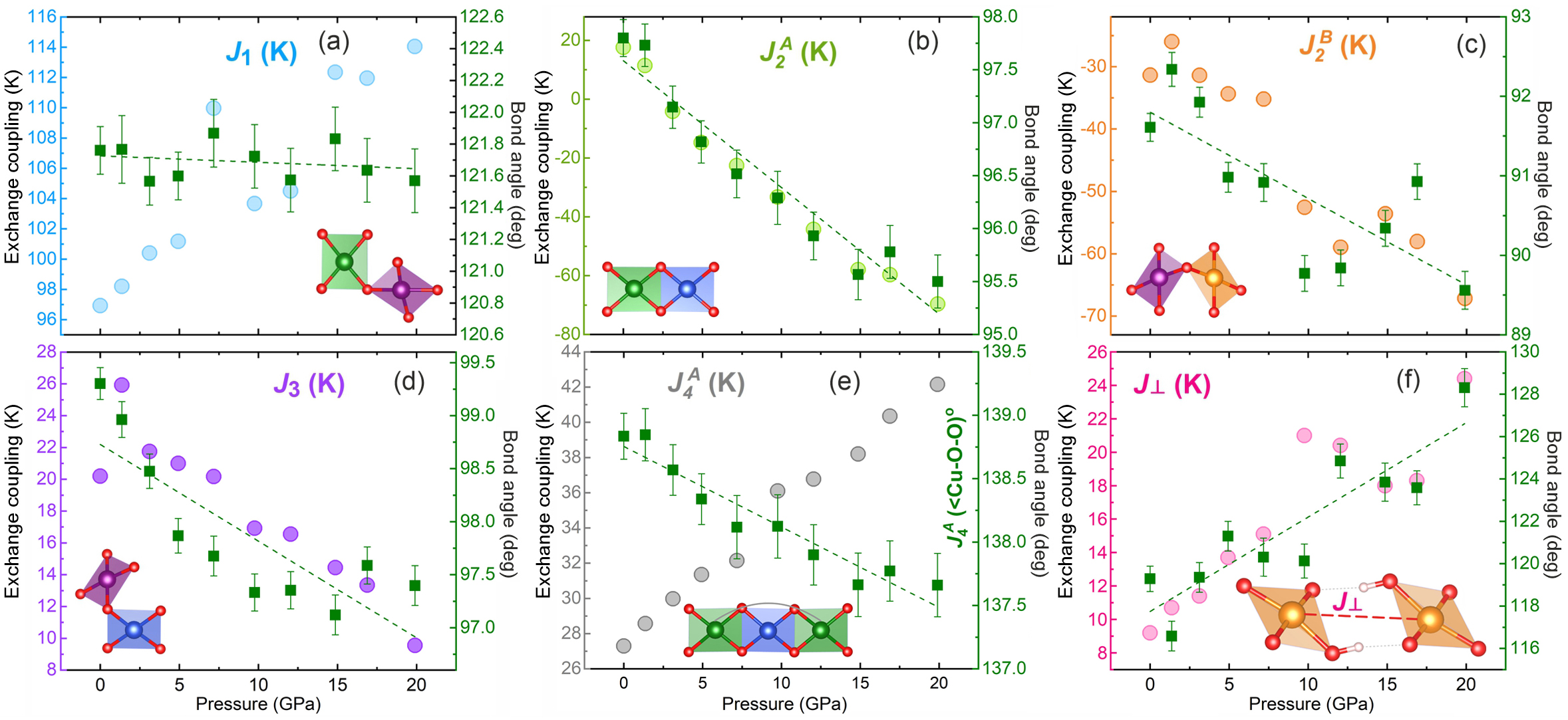}
    \caption{Pressure-dependent evolution of the averaged magnetic couplings $J_1$ (a), $J_2^A$ (b), $J_2^B$ (c), $J_3$ (d), $J_4^A$ (e), $J_{\perp}$ (f) and the corresponding structural parameters (bond angles) in brochantite: Cu--O--Cu in the case of $J_1-J_3$ and Cu--O--O in the case of $J_4^A$ and $J_{\perp}$ (see also Table~\ref{tab:exchange}). Dashed green lines show the trends in the bond angles as a guide for the eye. The inserts show the connectivity pathways of the copper polyhedra for the corresponding magnetic couplings.}
    \label{fig5}
\end{figure*}

We now calculate exchange couplings as a function of pressure using experimental structural data and allowing only hydrogen positions to relax. Despite some scatter in the individual $J_i$ values, this approach leads to very robust trends for the pressure evolution of the exchange parameters (Fig.~\ref{fig5}). The largest effect is seen in the case of $J_2^A$ that changes sign from AFM at ambient pressure to FM above 3\,GPa. Its counterpart $J_2^B$ also becomes more FM, but at a much smaller rate. Both trends arise from the reduction in the Cu--O--Cu bond angles in agreement with the Goodenough-Kanamori-Anderson rules. The twice larger value of $dJ/dP$ in the case of $J_2^A$ can be traced back to the stronger FM direct exchange in the edge-shared geometry.

A similar trend is seen in the coupling $J_3$ that becomes less AFM as its bond angle decreases. On the other hand, we observe a weak increase in the intrachain coupling $J_1$, whereas its bond angle remains unchanged within the accuracy of our measurement. This weak increase in $J_1$ is then probably related to more subtle effects beyond the nearest oxygen atoms, for example, to the changes in the position of the O--H bond, which is known to modulate the strength of superexchange via the oxygen atom~\cite{lebernegg2013b}.

Coming to the long-range couplings, we note that the interlayer exchange $J_{\perp}$ increases with pressure (Fig.~\ref{fig5}f). This trend is likely rooted in the increasing Cu--O--O angle that indicates a gradual evolution toward the linear Cu--O--O--Cu geometry that should be most favorable for the long-range superexchange. Interestingly, $J_4^A$ shows exactly the opposite trend, namely, this coupling increases upon compression, whereas the corresponding Cu--O--O angle decreases. We note, however, that the change in the angle between 0 and 20\,GPa is slightly above $-1$\,deg in the case of $J_4^A$ vs. $+10$\,deg in the case of $J_{\perp}$. It is then plausible that effects beyond the bond angle, in particular, the reduction in the O--O distance upon the shrinkage of the CuO$_4$ plaquettes (Fig.~\ref{fig4}) result in the enhancement of $J_4^A$ under pressure. 

The largest pressure-induced changes (Table~\ref{tab:djdp}) are observed for the couplings $J_2^A$ and $J_2^B$ with the Cu--Cu distances of about 3.0\,\r A and the significant contribution of direct exchange. All other couplings show moderate $dJ_i/dP$ of less than 1.0\,K/GPa regardless of the exact geometry. Interestingly, the couplings mediated by the \mbox{Cu--O--Cu} short-range superexchange and \mbox{Cu--O--O--Cu} long-range superexchange mechanisms show a similar tunability by pressure.

\begin{table}
\caption{\label{tab:djdp}
Pressure dependence of the exchange couplings in brochantite, $dJ_i/dP$ (in K/GPa). The values are obtained from the linear fits of the data shown in Fig.~\ref{fig5}.
}
\begin{ruledtabular}
\begin{tabular}{cccccc}
$J_1$ & $J_2^A$ & $J_2^B$ & $J_3$ & $J_4^A$ & $J_{\perp}$ \smallskip\\
0.8 & $-4.4$ & $-2.0$ & $-0.7$ & 0.8 & 0.7 \\ 
\end{tabular}
\end{ruledtabular}
\end{table}

\begin{figure*}[t]
  \centering
  \includegraphics[width=1\textwidth]{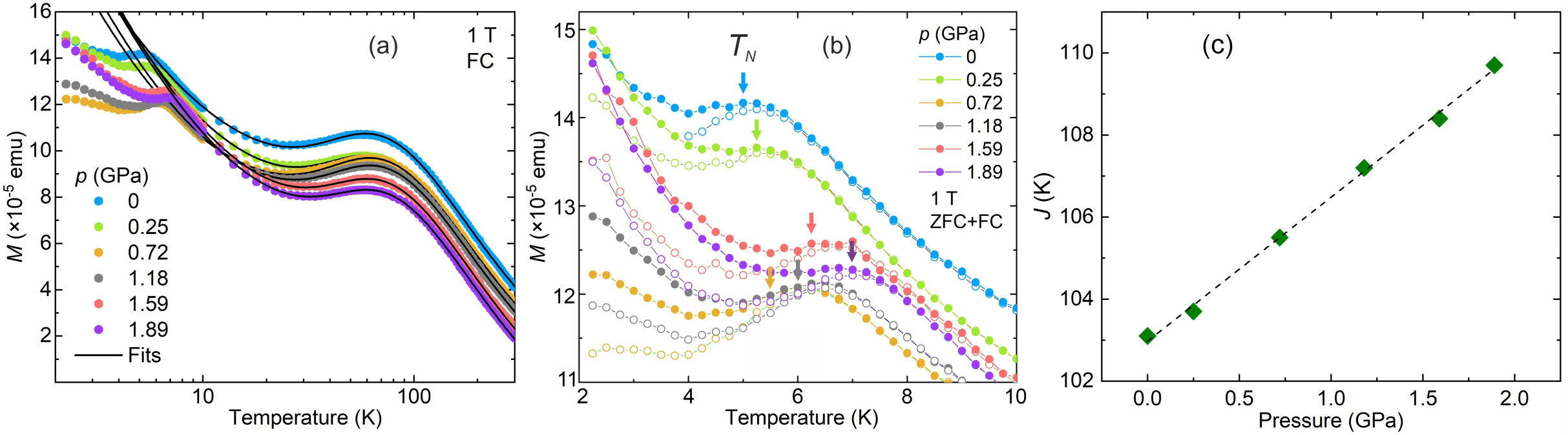}
    \caption{(a) Temperature-dependent dc magnetization of brochantite in a magnetic field of 1\,T at different pressures. The solid lines show the fits with Eq.~\eqref{eq:chi}. (b) Low-temperature part of magnetic susceptibility measured under field-cooled and zero-field-cooled conditions. The arrows label the bifurcation points and highlight the increase in $T_N$. (c) Pressure-dependent evolution of the exchange coupling $J$ extracted from the susceptibility fits.}
    \label{fig6}
\end{figure*}

Experimental magnetic susceptibility of brochantite measured under pressure is shown in Fig.~\ref{fig6}. The data are restricted to 1.9\,GPa because of the weak signal that would not be detectable in a smaller gasket required for reaching higher pressures. This narrow pressure interval is in fact already sufficient for tracing the pressure evolution of brochantite magnetism. The broad susceptibility maximum around 60\,K shifts toward higher temperatures on compression (Fig.~\ref{fig6}a). Fits with Eq.~\eqref{eq:chi} show a systematic increase in the intrachain exchange coupling $J$ with the slope of $3.2\pm 0.3$\,K/GPa (Fig.~\ref{fig6}c). This trend mirrors the enhancement of the intrachain coupling $J_1$, although that coupling increases much slower under pressure (Table~\ref{tab:djdp}). The larger $dJ/dP$ seen experimentally can be explained by the fact that the AFM order along the zigzag spin chains is stabilized not only by AFM $J_1$, but also by FM $J_2^A$ and $J_2^B$. Considering $J\simeq J_1-(J_2^A+J_2^B)/2$ and using the values from Table~\ref{tab:djdp}, we estimate $dJ/dP=4.0$\,K/GPa in a better agreement with the experimental value of 3.2\,K/GPa. Remarkably, the rise in the FM interactions $J_2$ makes the dominant contribution to the pressure-induced enhancement of the AFM intrachain coupling $J$.

We also used the susceptibility data to track the pressure dependence of $T_N$. While the kink at $T_N$ is largely smeared out in these measurements, the transition is visible by the weak splitting of the susceptibility measured under field-cooled and zero-field-cooled conditions (Fig.~\ref{fig6}b). The transition shifts from 5.0\,K at ambient pressure to around 7.0\,K at 1.9\,GPa. This increase in $T_N$ with $dT_N/dP=0.96$\,$\pm 0.10$\,K/GPa is comparable in magnitude to the pressure-induced changes in $J_3$ and $J_{\perp}$ (Table~\ref{tab:djdp}), in agreement with the fact that 3D magnetic order in spin-chain systems is largely determined by the interchain couplings. The frustration of brochantite should be somewhat reduced under pressure because $J_2^A$ becomes FM. However, the frustration due to $J_4^A$ persists.



\section{Discussion and Summary}

Although many aspects of the brochantite magnetism can be understood within the simple model of AFM spin chains, their formation in the complex layered crystal structure is far from obvious. Our analysis shows that such zigzag spin chains arise from the longer Cu--Cu distances of $3.5-3.6$\,\r A spanning the structural chains A and B. However, the AFM coupling $J_1$ along the spin chains is by far not the only relevant interaction in this material. Pressure dependence of the effective intrachain coupling $J$ clearly shows that the AFM spin chains are additionally stabilized by the couplings $J_2$, with $J_2^B$ being FM already at ambient pressure, whereas $J_2^A$ is initially AFM and becomes FM upon compression. Several long-range couplings frustrate the collinear magnetic structure of brochantite and result in a quite low $T_N/J\simeq 0.05-0.06$, despite the very short distances and sizable exchange couplings between the adjacent spin chains. 

\begin{table}
\caption{\label{tab:comparison}
Comparison of the structural chains A and B in several Cu$^{2+}$ minerals. The lattice parameter $b$ (in \r A) describes the periodicity of the chains, whereas $J_2^A$ and $J_2^B$ (in K) are the respective nearest-neighbor couplings.
}
\begin{ruledtabular}
\begin{tabular}{ccccrc}
 mineral     & composition                       &    $b$     & $J_2^A$  & $J_2^B$ & Ref.\smallskip\\
 brochantite & Cu$_4$SO$_4$(OH)$_6$              &   6.02     &  18         &  $-32$     & \\
 langite     & Cu$_4$SO$_4$(OH)$_6\cdot 2$H$_2$O &   6.03     &  24         &  $-49$     & \cite{lebernegg2016} \\
 rouaite     & Cu$_2$(OH)$_3$NO$_3$              &   6.07     &  60         &    3       & \cite{yuan2022}  \\
 antlerite    & Cu$_3$SO$_4$(OH)$_4$             &   6.08     &  42         &  $-21$     & \cite{kulbakov2022a} \\
 botallackite & Cu$_2$(OH)$_3$Br                 &   6.16     &  62         &  $-19$     & \cite{zhang2020} \\ 
\end{tabular}
\end{ruledtabular}
\end{table}

The largest pressure dependence of about $-4$\,K/GPa is observed for the coupling $J_2^A$ that corresponds to the edge-shared CuO$_4$ plaquettes and hinges upon a delicate interplay of AFM superexchange and FM direct exchange. Pressure reduces the former and enhances the latter by reducing the Cu--O--Cu bond angles and the respective Cu--Cu distance. These trends explain the crossover from AFM to FM $J_2^A$ under pressure. Several other magnetic couplings show an opposite evolution, and here it is remarkable that the same pressure dependence of about $+1$\,K/GPa is observed for $J_1$ mediated by Cu--O--Cu superexchange and $J_4^A$ or $J_{\perp}$ that involve superexchange via two oxygen atoms. Consequently, brochantite remains frustrated under pressure, although its N\'eel temperature increases as the couplings along the spin chains ($J_1$, $J_2^A$, $J_2^B$) and between the layers ($J_{\perp}$) are enhanced.

The pressure dependence of the exchange couplings, especially $J_2^A$ and $J_2^B$, can be used to rationalize the behavior of several frustrated Cu$^{2+}$ minerals. These structural siblings of brochantite are listed in Table~\ref{tab:comparison} along with their exchange couplings and the lattice parameter $b$ taken along the structural chains. The AFM $J_2^A$ and FM $J_2^B$ are clearly distinguished. Moreover, increasing the lattice parameter (negative pressure) renders $J_2^A$ more AFM and $J_2^B$ less FM, in agreement with the pressure dependence observed in our work (Fig.~\ref{fig5}). The same trend can be seen in the experimental magnetic structures where chains A develop collinear AFM order in botallackite~\cite{zhang2020} and antlerite~\cite{kulbakov2022a} and a twisted AFM-like order in rouaite~\cite{peets2024}. By contrast, chains B develop collinear FM order in botallackite~\cite{zhang2020} and rouaite~\cite{peets2024} and a twisted FM-like order in antlerite~\cite{kulbakov2022a}. The magnetic ground state should of course also depend on the couplings $J_1$ and $J_3$ between the structural chains. Here, brochantite seems to be unique in that it features $J_1\gg J_3$, likely because of the buckled layer geometry (Fig.~\ref{fig1}e) that allows increased Cu--Cu distances and Cu--O--Cu bond angles for $J_1$. Other materials listed in Table~\ref{tab:comparison} are expected to show comparable $J_1$ and $J_3$~\cite{zhang2020,kulbakov2022a}, thus being further away from the 1D limit manifested by brochantite. 

Finally, we note that hydrostatic pressure and uniaxial strain are both powerful tools for tailoring quantum magnetism of Cu$^{2+}$ compounds~\cite{wang2023,fogh2024,chatterjee2025}. The microscopic evaluation of the pressure and strain effects is often hindered by the lack of accurate structural data, because complex sample environment restricts the accessible reciprocal space and renders the full determination of the crystal structure impossible. Our study shows that even with the limited part of the reciprocal space accessible in a diamond anvil cell, the Cu and O positions in the complex structure of brochantite can be resolved as a function of pressure with the accuracy sufficient for the quantitative analysis of magnetic couplings, which are known to be highly sensitive to even minor alterations of the structural geometry. This sets a benchmark for future studies of pressure and strain effects in cuprates without relying on \textit{ab initio} structural relaxations, such that the structural information can be obtained from the experiment only. The pressure dependencies determined in our work for different types of exchange couplings (Table~\ref{tab:djdp} and Fig.~\ref{fig5}) can be directly applied to other Cu$^{2+}$ materials for a preliminary assessment of their evolution under pressure and strain.


\acknowledgments
We thank Artem S. Borisov (Kiel University) for generously providing a natural brochantite specimen from his private collection. We are also grateful to Prof. H. Krautscheid (Leipzig University) for access to the single-crystal X-ray diffraction equipment, and to Martin Börner (Leipzig University) for his help in organizing the single-crystal XRD experiment at ambient pressure. AT thanks Oleg Janson and Darren Peets for fruitful discussions. We acknowledge DESY (Hamburg, Germany), a member of the Helmholtz Association HGF, for the provision of experimental facilities. Parts of this research were carried out at the P02.2 beamline of PETRA III. Beamtime was allocated for the proposal I-20230876. We also acknowledge ESRF (Grenoble, France) for providing the synchrotron beamtime. The work in Leipzig was funded by the Deutsche Forschungsgemeinschaft (DFG, German Research Foundation) -- TRR 360 -- 492547816 (subproject B1). 
\smallskip

\textit{Data availability.} Research data associated with this study can be found in Ref.~\cite{data}.


%

\end{document}